\documentclass{amsart}

\usepackage{graphicx}
\usepackage{amsmath}
\usepackage{amssymb}

\usepackage{color}

\definecolor{red}{rgb}{1,0,0}
\def\red{\color{red}}
\definecolor{green}{rgb}{0,1,0}

\definecolor{blue}{rgb}{0,0,1}

\definecolor{black}{rgb}{0,0,0}

\definecolor{grey}{rgb}{0.333,0.333,0.333}

\definecolor{white}{rgb}{1,1,1}

\newtheorem{thm}{Theorem}

\def\QQ{{\mathbb Q}}
\def\PP{{\mathbb P}}
\def\RR{{\mathbb R}}

\def\ZZ{{\mathbb Z}}

\def\vece{{\text{\boldmath$e$}}}

\def\vecn{{\text{\boldmath$n$}}}
\def\vecq{{\text{\boldmath$q$}}}
\def\vecQ{{\text{\boldmath$Q$}}}

\def\vecS{{\text{\boldmath$S$}}}
\def\uvecS{\widehat{\text{\boldmath$S$}}}

\def\vecv{{\text{\boldmath$v$}}}
\def\vecV{{\text{\boldmath$V$}}}

\def\vecx{{\text{\boldmath$x$}}}

\def\vecy{{\text{\boldmath$y$}}}

\def\vecalf{{\text{\boldmath$\alpha$}}}

\def\vecxi{{\text{\boldmath$\xi$}}}

\def\vecnull{{\text{\boldmath$0$}}}

\def\scrA{{\mathcal A}}
\def\scrB{{\mathcal B}}

\def\scrD{{\mathcal D}}

\def\scrK{{\mathcal K}}
\def\scrL{{\mathcal L}}

\def\scrZ{{\mathcal Z}}

\def\e{\mathrm{e}}

\def\diag{\operatorname{diag}}

\def\id{\operatorname{id}}

\def\C{\operatorname{C{}}}

\def\L{\operatorname{L{}}}

\def\S{\operatorname{S{}}}

\def\SL{\operatorname{SL}}
\def\ASL{\operatorname{ASL}}
\def\SO{\operatorname{SO}}

\def\T{\operatorname{T{}}}

\def\ASLASL{\ASL(d,\ZZ)\backslash\ASL(d,\RR)}
\def\ASLZ{\ASL(d,\ZZ)}
\def\ASLR{\ASL(d,\RR)}
\def\SLSL{\SL(d,\ZZ)\backslash\SL(d,\RR)}
\def\SLZ{\SL(d,\ZZ)}
\def\SLR{\SL(d,\RR)}

\def\trans{\,^\mathrm{t}\!}

\begin{document}

\title{\uppercase{Kinetic transport in crystals}}
\author{\uppercase{Jens Marklof}}
\address{School of Mathematics, University of Bristol,
Bristol BS8 1TW, U.K.}
\email{j.marklof@bristol.ac.uk}
\thanks{Submitted to the Proceedings of the XVIth International Congress of Mathematical Physics, Prague 2009}

\begin{abstract}
One of the central challenges in kinetic theory is the derivation of macroscopic evolution equations---describing, for example, the dynamics of an electron gas---from the underlying fundamental microscopic laws of classical or quantum mechanics. An iconic mathematical model in this research area is the Lorentz gas, which describes an ensemble of non-interacting point particles in an infinite array of spherical scatterers. In the case of a disordered scatterer configuration, the classical results by Gallavotti, Spohn and Boldrighini-Bunimovich-Sinai show that the time evolution of a macroscopic particle cloud is governed, in the limit of small scatterer density (Boltzmann-Grad limit), by the linear Boltzmann equation. In this lecture I will discuss the recent discovery that for a periodic configuration of scatterers the linear Boltzmann equation fails, and the random flight process that emerges in the Boltzmann-Grad limit is substantially more complicated. The key ingredient in the description of the limiting stochastic process is the renormalization dynamics on the space of lattices, a powerful technique that has recently been successfully applied also to other open problems in mathematical physics, including KAM theory and quantum chaos. This lecture is based on joint work with Andreas Str\"ombergsson, Uppsala.
\end{abstract}

\maketitle


\section{Introduction \label{secIntro}}

An important cornerstone in mathematical physics is the problem of deriving macroscopic evolution equations from first principles, i.e., the microscopic laws of motion governed by quantum theory, or (to simplify) Newton's laws of classical mechanics. The subject has its origin in Boltzmann's revolutionary vision formulated more than a century ago, and it is perhaps surprising that still today there is no complete understanding of his most fundamental model, a dilute gas of hard spheres: Lanford's seminal work \cite{Lanford74} establishes the validity of the Boltzmann equation only for times that are a fraction of the mean collision time. Even simpler models, such as a gas of non-interacting particles in a fixed array of scatterers studied first by Lorentz in 1905 \cite{Lorentz} (the {\em Lorentz gas}), are difficult to analyze and lead to new and unexpected macroscopic phenomena. This lecture will focus on the Lorentz gas and report on recent joint work with Andreas Str\"ombergsson \cite{partI}, \cite{partII}, \cite{partIII}, \cite{partIV} in the case of a crystal with periodic scatterer configuration. The exciting aspect of our findings is that the kinetic transport equations that emerge in the limit of small scatter size (the {\em Boltzmann-Grad limit}) are new, and distinctly different from the answer for disordered scatterers.

\begin{figure}
\begin{center}
\includegraphics[width=0.4\textwidth]{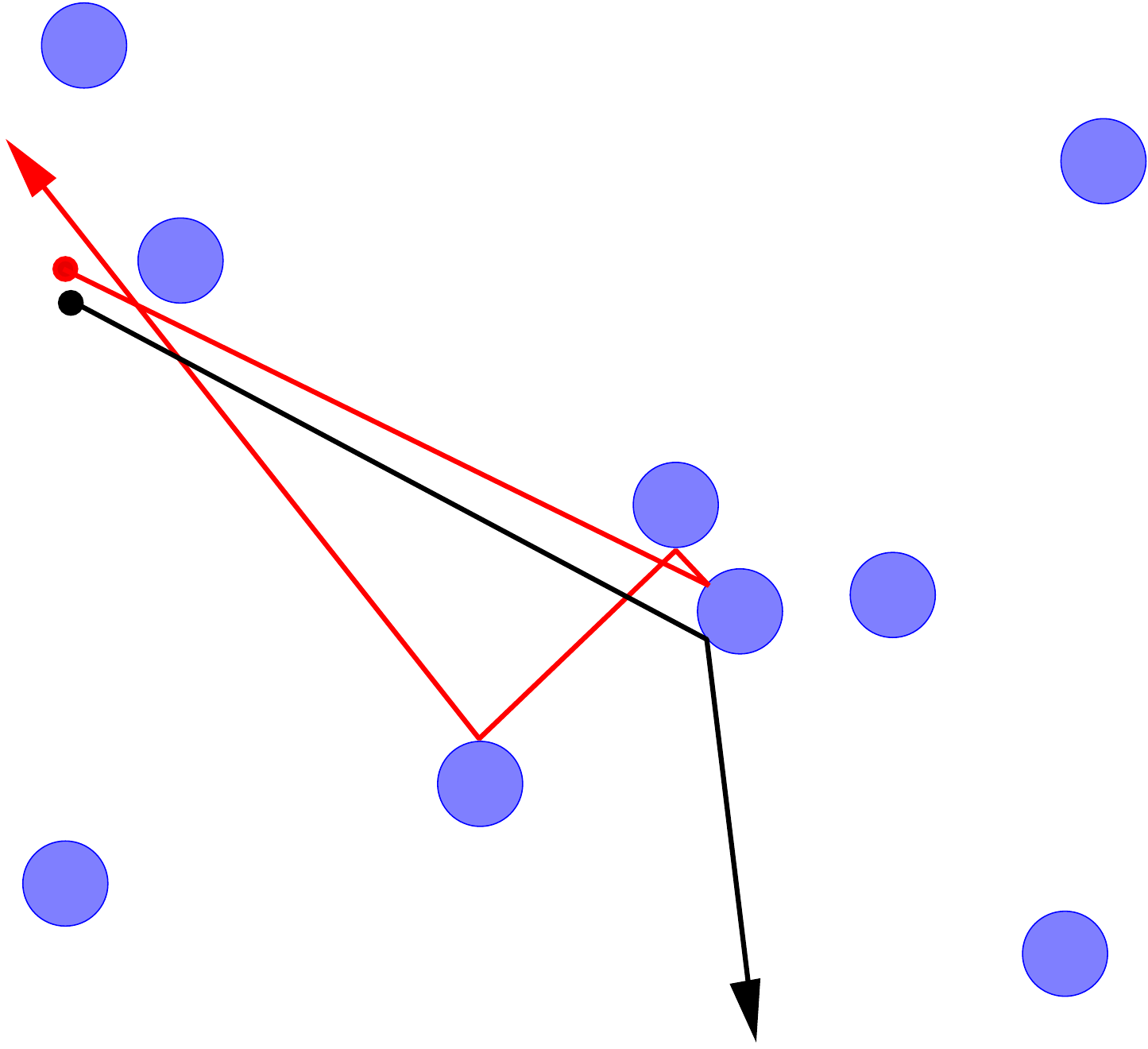}
\end{center}
\caption{The Lorentz gas for a random scatterer configuration. \label{figL}}
\end{figure}

In his original work, Boltzmann considered the case of a dilute gas of hard spheres with elastic collisions; {\em dilute} refers to the limit of small particle density, i.e., the ratio of particle radius and mean separation tends to zero. In the case of a gas of non-interacting particles in an array of fixed scatterers---the Lorentz gas---the dynamics can be reduced to a one-particle motion (Fig.~\ref{figL}). We denote by $\vecq(t)\in\RR^d$ the position and by $\vecv(t)\in\RR^d$ the velocity of our particle. The correct kinetic limit is then obtained by taking the scatterer radius $\rho\to 0$. A simple dimensional argument shows that in this limit the mean free path length should scale like $\rho^{-(d-1)}$, i.e., the inverse of the total scattering cross section of an individual scatterer. This suggests to rescale the length units by introducing the macroscopic coordinates
\begin{equation}\label{macQV}
	\big(\vecQ(t),\vecV(t)\big) = \big(\rho^{d-1} \vecq(\rho^{-(d-1)} t),\vecv(\rho^{-(d-1)} t)\big) .
\end{equation}
This rescaling of length and time is commonly referred to as the {\em Boltzmann-Grad scaling}, and the corresponding limit $\rho\to 0$ as the {\em Boltzmann-Grad limit}.
The time evolution of a particle with initial data $(\vecQ,\vecV)$ is then described by the billiard flow
\begin{equation}\label{LP}
	(\vecQ(t),\vecV(t))=\varPhi_\rho^t(\vecQ,\vecV) .
\end{equation}
Since the speed of our particle is a constant of motion we may assume without loss of generality that $\|\vecV\|=1$. For notational reasons it is convenient to extend the dynamics to the inside of each scatterer trivially, i.e., set $\varPhi_\rho^t=\id$ whenever $\vecQ$ is inside the scatterer. That is, the relevant phase space is now the unit tangent bundle of $\RR^d$, which will be denoted by $\T^1(\RR^d)$. 

The time evolution of an initial particle density $f\in\L^1(\T^1(\RR^d))$ is 
\begin{equation}
	f_t= L_\rho^t f
\end{equation}
where $L_\rho^t$ is the Liouville operator defined by
\begin{equation}
	[L_\rho^t f](\vecQ,\vecV) := f\big(\varPhi_\rho^{-t}(\vecQ,\vecV)\big) .
\end{equation}
Following Boltzmann's arguments, Lorentz concluded in his 1905 paper that the macroscopic time evolution of a particle cloud should, in the limit $\rho\to 0$, be governed by the {\em linear Boltzmann equation} (today also referred to as {\em kinetic Lorentz equation}),
\begin{equation}
		\bigg[ \frac{\partial}{\partial t} + \vecV\cdot\nabla_\vecQ \bigg] f_t(\vecQ,\vecV)= \int_{\S_1^{d-1}}  \big[ f_t(\vecQ,\vecV_0)-f_t(\vecQ,\vecV)\big] \sigma(\vecV_0,\vecV) d\vecV_0,
\end{equation}
where the collision kernel $\sigma(\vecV_0,\vecV)$ is the differential cross section of the individual scatterer. In the case of elastic scattering at hard spheres we have $\sigma(\vecV_0,\vecV)=\frac14 \|\vecV_0-\vecV\|^{3-d}$. The linear Boltzmann equation describes a random flight process, where a particle moves freely with constant velocity $\vecV_0$ for time $t\geq \xi$ with probability $\e^{-\nu_d \xi}$, where $\nu_d$ is the volume of the $(d-1)$-dimensional unit ball (the total scattering cross section), and is then scattered to velocity $\vecV$ with probability $\sigma(\vecV_0,\vecV)$, again flies with constant velocity $\vecV$ for time $t\geq \xi$ with probability $\e^{-\nu_d \xi}$, and so on. The crucial observation is that each scattering event is independent of the previous one. Thus the process that generates the paths of our random flight is Markovian. As we will see, this is different in the case of a periodic scatterer configuration.

The validity of the linear Boltzmann equation was first established rigorously for a random, Poisson distributed scatterer configuration by Gallavotti \cite{Gallavotti69}. His results were generalized by Spohn \cite{Spohn78} to more general random scatterer configurations and scattering potentials. In 1983, Boldrighini, Bunimovich and Sinai \cite{Boldrighini83} proved convergence for almost every scatterer configuration drawn from a Poisson distribution. As Spohn's work shows, the details of the randomness of the scatterer positions is not so essential, and as long as there are no strong correlations, all of the above results should remain valid. 

The linear Boltzmann equation has numerous important applications, e.g., to neutron transport and radiative transfer, and it is thus natural to ask under which circumstances it may fail to provide an accurate description. 
	
\section{The periodic Lorentz gas}	
	
\begin{figure}
\begin{center}
\begin{minipage}{0.8\textwidth}
\unitlength0.1\textwidth
\begin{picture}(10,6)(0,0)
\put(0,0){\includegraphics[width=\textwidth]{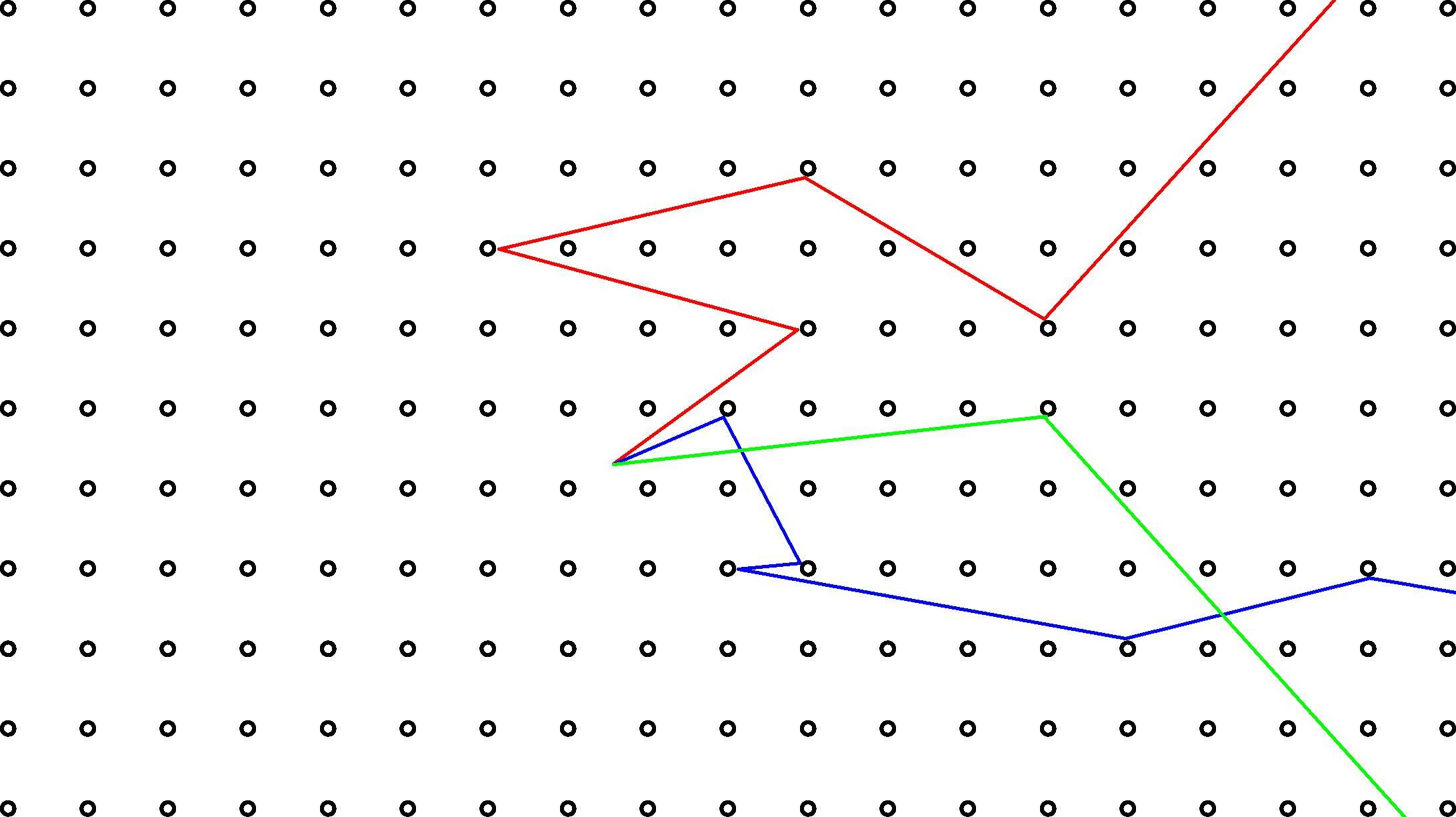}}
\put(4.5,3.0){$\red \vecS_1$} \put(3.7,3.5){\red $\vecS_2$}
\put(4.5,4.3){$\red\vecS_3$} \put(6.1,3.6){$\red\vecS_4$} \put(8,4.8){\red $\vecS_5$}
\end{picture}
\end{minipage}
\end{center}
\caption{The Lorentz gas for a periodic scatterer configuration, with three distinct particle trajectories. \label{figPL}}
\end{figure}
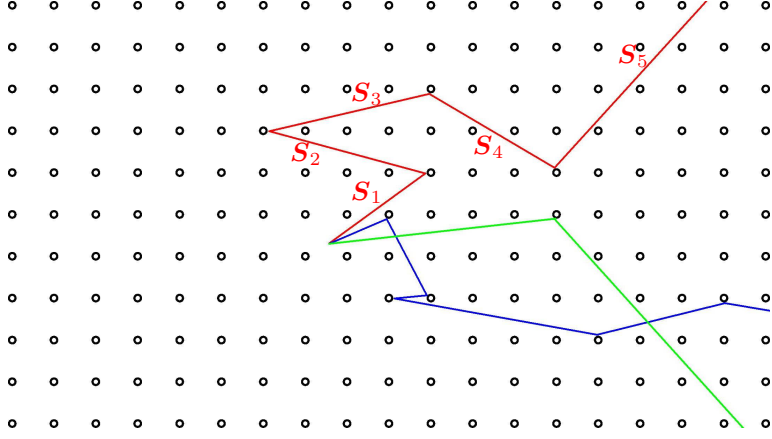

Given a euclidean lattice $\scrL\subset\RR^d$ of covolume one (e.g., $\scrL=\ZZ^d$) the {\em periodic Lorentz gas} is defined as the dynamics of a cloud of non-interacting point particles in an array of identical scatterers that are placed at the vertices of the lattice $\scrL$ (Fig.~\ref{figPL}). The periodic Lorentz gas has served as a fundamental model in the understanding of chaotic diffusion, which emerges in the long-time limit with {\em fixed} scatterer size. In their pioneering work, Bunimovich and Sinai \cite{Bunimovich80} proved a central limit theorem for the long-time dynamics of a particle cloud in two dimensions. That is, the long-time evolution converges, in the appropriate scaling limit, to a solution of the heat equation. More refined statistical properties that show that the dynamics in fact converges to Brownian motion, have recently been established in the work of Melbourne and Nichol \cite{Melbourne05}, \cite{Melbourne07}, and the recent paper by Dolgopyat, Szasz and Varju \cite{dolgopyat}. All of the above results assume that the Lorentz gas has {\em finite horizon}, i.e., there are no infinitely long free flight paths. Without this assumption, a central limit theorem still holds, but the diffusion constant is no longer linear in time $t$ but diverges as $t\log t$. This was observed by Bleher \cite{Bleher92} and recently established rigorously by Szasz and Varju \cite{Szasz07}. It is interesting that none of these results have so far been extended to dimension $d\geq 3$. The arguments given by Chernov \cite{Chernov94} and by Balint and Toth \cite{Balint07} require non-trivial hypotheses that are difficult to establish. 

We now return to the question of the existence of kinetic transport equations in the Boltzmann-Grad limit discussed in the previous section. Recall that we are interested in the dynamics of a particle cloud
\begin{equation}
	f_t(\vecQ,\vecV)=[L^t_\rho f](\vecQ,\vecV) = f\big(\varPhi_\rho^{-t}(\vecQ,\vecV)\big) 
\end{equation}
in the macroscopic coordinates \eqref{macQV}. 
The estimates by Bourgain, Golse and Wennberg \cite{Bourgain98}, \cite{Golse00} on the distribution of free path lengths already imply that the linear Boltzmann equation does not hold in the periodic set-up; this was pointed out recently by Golse \cite{Golse07}. 

The first key result of our joint work with Str\"ombergsson is the proof of the existence of a limiting random process for the periodic Lorentz gas \cite[Section 1]{partII}.

\begin{thm}\label{thmA}
Fix a euclidean lattice $\scrL$. For every $t>0$ there exists a linear operator $L^t:\L^1(\T^1(\RR^d))\to\L^1(\T^1(\RR^d))$, such that for every $f\in\L^1(\T^1(\RR^d))$ and
any set $\scrA\subset\T^1(\RR^d)$ with boundary of Lebesgue measure zero,
\begin{equation}
	\lim_{\rho\to 0} \int_{\scrA} [L^t_\rho f](\vecQ,\vecV)\, d\vecQ\, d\vecV 
	= \int_{\scrA} [L^t f](\vecQ,\vecV) \, d\vecQ\, d\vecV .
\end{equation}
\end{thm}
	
\begin{figure}
\begin{center}
\begin{minipage}{0.49\textwidth}
\unitlength0.1\textwidth
\begin{picture}(10,8)(0,0)
\put(0.5,1){\includegraphics[width=0.9\textwidth]{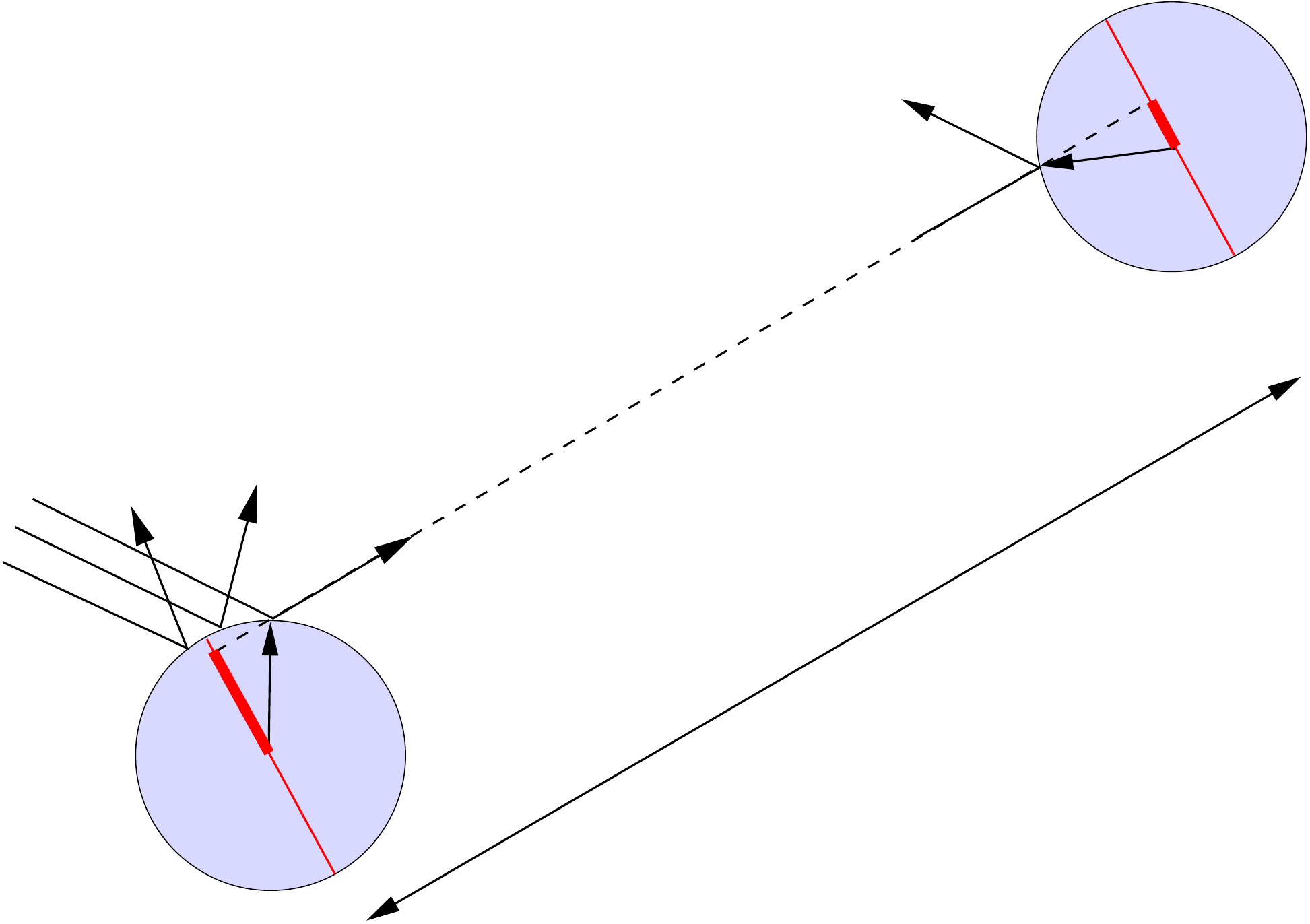}}
\put(0.1,3.2){$\vecV_0$}
\put(2.3,3.3){$\vecV$} 
\put(6.6,5.8){$\vecV$} 
\put(1.6,2.4){$\rho s$}
\put(6,2.5){$\rho^{-(d-1)}\xi$}
\put(8.6,6.6){$\rho b$}
\put(6.8,6.8){$\vecV_+$} 
\end{picture}
\end{minipage}
\end{center}
\caption{Two consecutive collisions in the Lorentz gas. \label{figTP}}
\end{figure}
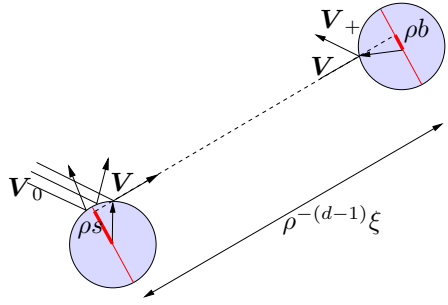

The operator $L^t$ thus describes the macroscopic diffusion of the Lorentz gas in the Boltzmann-Grad limit $\rho\to 0$. As we shall see however, the family $\{L^t\}_{t\geq 0}$ does {\em not} form a semigroup, i.e.,
\begin{equation}
	L^s L^t \neq L^{s+t}. 
\end{equation}
This is perhaps surprising since $\{L_\rho^t\}_{t\geq 0}$ is indeed a semigroup for every fixed $\rho>0$. What is more, in the case of the random scatterer configuration the corresponding limiting operators $L^t$ also form a semigroup---after all, $f_t:=L^t f$ is a solution of the linear Boltzmann equation. The reason for the failure of the semigroup property in the periodic setting stems from additional correlations in the lattice, which are lost in the macroscopic scaling limit. To keep track of this data, we consider extended phase space coordinates 
$(\vecQ,\vecV,\xi,\vecV_+)$
where $(\vecQ,\vecV)\in\T^1(\RR^d)$ is the usual position and momentum, $\xi\in\RR_+$ the flight time until the next collision, and $\vecV_+\in\S_1^{d-1}$ the velocity after the next collision. On the microscopic level, the system is now over-determined ($\xi$ and $\vecV_+$ are functions of $\vecQ$ and $\vecV$), but on the macroscopic scale the extra variables are needed. We prove in \cite[Section 6]{partII} that the particle density $f_t(\vecQ,\vecV,\xi,\vecV_+)$ indeed satisfies a {\em generalized} linear Boltzmann equation
\begin{equation}\label{glB}
	\bigg[ \frac{\partial}{\partial t} + \vecV\cdot\nabla_\vecQ - \frac{\partial}{\partial\xi} \bigg] f_t(\vecQ,\vecV,\xi,\vecV_+) 
	= \int_{\S_1^{d-1}}  f_t(\vecQ,\vecV_0,0,\vecV) 
p_{\vecnull}(\vecV_0,\vecV,\xi,\vecV_+) \,
d\vecV_0 .
\end{equation} 
The left hand side again corresponds to free transport (note that $\xi$ is decreasing linearly with $t$). The right hand side involves a new collision kernel $p_{\vecnull}(\vecV_0,\vecV,\xi,\vecV_+)$, given by
\begin{equation}
	p_{\vecnull}(\vecV_0,\vecV,\xi,\vecV_+) 
	=\sigma(\vecV,\vecV_+)\, \varPhi_\vecnull\big(\xi,b(\vecV,\vecV_+),
	-s(\vecV,\vecV_0)\big)
\end{equation}
where $\sigma(\vecV,\vecV_+)$ is the differential cross section and $\varPhi_\vecnull\big(\xi,b(\vecV,\vecV_+),-s(\vecV,\vecV_0)\big)$ the transition probability density to exit with parameter $s(\vecV,\vecV_0)$ and hit the next scatterer at time $\xi$ with impact parameter $b(\vecV,\vecV_+)$; cf.~Fig.~\ref{figTP}.

In dimension $d=2$ we have the following explicit formula for the transition probability \cite{partIII}:
\begin{equation}\label{Xp}	\varPhi_\vecnull(\xi,w,z)=\frac{6}{\pi^2}\Upsilon\Bigl(1+\frac{\xi^{-1}-\max(|w|,|z|)-1}{|w+z|}\Bigr)
\end{equation}
with
\begin{equation}
	\Upsilon(x)=
\begin{cases} 
0 & \text{if }x\leq 0\\
x & \text{if }0<x<1\\
1 & \text{if }1\leq x,
\end{cases}
\end{equation}
The same formula has recently been found independently by Caglioti and Golse \cite{Caglioti08} and by Bykovskii and Ustinov \cite{Bykovskii09}, using different methods based on continued fractions.
 
Our formulas for dimension $d\geq 3$ are not as explicit and substantially more involved, see \cite{partI},\cite{partII} for details, and \cite{partIV} for asymptotic tail estimates. The formulas imply in particular that the collision kernel $p_0$ (and thus the limiting process) is {\em independent} of the lattice $\scrL$ on which the scatterers are positioned. Hence any microscopic preference for certain directions completely disappears in the Boltzmann-Grad limit. 

The operators $L^t$ in Theorem \ref{thmA} we were originally interested in can be recovered by integrating over the auxiliary variables $\xi$ and $\vecV_+$,
\begin{equation}
	[L^t g](\vecQ,\vecV):= \int_0^\infty \int_{\S_1^{d-1}} f_t(\vecQ,\vecV,\xi,\vecV_+) \, d\vecV_+\, d\xi
\end{equation}
where $f_t(\vecQ,\vecV,\xi,\vecV_+)$ is a solution of the generalized linear Boltzmann equation subject to the initial condition 
\begin{equation}
	\lim_{t\to 0}f_t(\vecQ,\vecV,\xi,\vecV_+) = g(\vecQ,\vecV) p(\vecV,\xi,\vecV_+) 
\end{equation}
with
	\begin{equation}
	p(\vecV,\xi,\vecV_+) := \int_\xi^\infty \int_{\S_1^{d-1}} \sigma(\vecV_0,\vecV) p_{\vecnull}(\vecV_0,\vecV,\xi',\vecV_+)\, d\vecV_0\, d\xi';
	\end{equation}
the latter is a stationary solution of the generalized linear Boltzmann equation.
	
\section{Why ``a generalization'' of the linear Boltzmann equation?}	

The reason why \eqref{glB} is indeed a generalization of the linear Boltzmann equation is the following. As mentioned in the introduction, the linear Boltzmann equation corresponds to a random flight process where the time $\xi$ until the next collision has probability density $\nu_d\e^{-\nu_d\xi}$, where $\nu_d$ is the volume of the $(d-1)$-dimensional unit ball. Furthermore, the probability to exit with parameter $s(\vecV,\vecV_0)$ and hit the next scatterer with impact parameter $b(\vecV,\vecV_+)$ should be uncorrelated and independent of $\xi$. We have thus
\begin{equation}
	\varPhi_\vecnull\big(\xi,b(\vecV,\vecV_+),-s(\vecV,\vecV_0)\big)
	= \e^{-\nu_d \xi}.
\end{equation}
Substituting in the above the transition density for the random (rather than periodic) scatterer configuration, we obtain
\begin{equation}
p_{\vecnull}(\vecV_0,\vecV,\xi,\vecV_+) = \sigma(\vecV,\vecV_+) \, \e^{-\nu_d\xi} 
= p(\vecV,\xi,\vecV_+)
\end{equation}
and
\begin{equation}
f_t(\vecQ,\vecV,\xi,\vecV_+) = g_t(\vecQ,\vecV)\,\sigma(\vecV,\vecV_+) \,  \e^{-\nu_d\xi} .
\end{equation}
It is now straightforward to see that \eqref{glB} yields the classical linear Boltzmann equation for $g_t(\vecQ,\vecV)$.	
	
\section{Joint distribution of path segments}	
	
The following theorem is the central result in our investigation \cite{partII}. It shows that the limiting random flight process exists, and that the consecutive path segments $\vecS_1,\vecS_2,\ldots$ are generated by a Markov process with memory two. That is, the probability for the $n$th leg of our path only depends on the previous two path segments.

We will use the notation $\uvecS:=\|\vecS\|^{-1}\vecS$. We set
\begin{equation}
\scrB_n:=\bigl\{(\vecS_1,\ldots,\vecS_n)\in(\RR^{d}\setminus\{\vecnull\})^n:\; \uvecS_{j+1}\neq\uvecS_{j}\; (j=1,\ldots,n-1) \bigr\} .
\end{equation}

\begin{thm}\label{secThmMacro}
Fix a lattice $\scrL$ and let $\Lambda$ be a Borel probability measure on $\T^1(\RR^d)$ which is absolutely continuous with respect to Lebesgue measure. Then, for each $n\in\ZZ_{>0}$, and for any set $\scrA\subset \RR^d\times\RR^{nd}$ with boundary of Lebesgue measure zero,
\begin{multline} \label{secThm-eq-macro}
	\lim_{\rho\to 0}\Lambda\big( \big\{ (\vecQ_0,\vecV_0)\in \T^1(\rho^{d-1}\scrK_\rho) : (\vecQ_0,\vecS_1(\vecQ_0,\vecV_0;\rho),\ldots,\vecS_n(\vecQ_0,\vecV_0;\rho)) \in \scrA \big\} \big) \\
= \int_{\scrA} P^{(n)}(\vecS_1,\ldots,\vecS_n)\, \Lambda'\big(\vecQ_0,\uvecS_1\big)\, d\vecQ_0\, d\vecS_1 \cdots d\vecS_n ,
\end{multline}
and where $\Lambda'$ is the Radon-Nikodym derivative of
$\Lambda$ with respect to Lebesgue measure.
Furthermore, there is a function $\Psi:\scrB_3\to\RR_{\geq 0}$ 
such that
\begin{equation} \label{jointlimdens}
	P^{(n)}(\vecS_1,\ldots,\vecS_n) = P^{(2)}(\vecS_1,\vecS_2)
	\prod_{j=3}^n \Psi(\vecS_{j-2},\vecS_{j-1},\vecS_j) 
\end{equation}
for all $n\geq 3$ and all $(\vecS_1,\ldots,\vecS_n)\in\scrB_n$.
\end{thm}

We in fact also prove a refined version of this theorem, where the initial position $\vecQ_0$ is not random but fixed (on the microscopic scale); see \cite[Theorem 1.1]{partII}. Furthermore, the limiting distributions $P^{(n)}$ and $\Psi$ are {\em independent} of $\scrL$ and $\Lambda$.

We define the probability measure corresponding to \eqref{jointlimdens} by 
\begin{equation} \label{muLambdandef}
	\mu_{\Lambda}^{(n)}(\scrA):= \int_{\scrA} P^{(n)}(\vecS_1,\ldots,\vecS_n)\, \Lambda'\big(\vecQ_0,\uvecS_1\big)\, d\vecQ_0\, d\vecS_1\cdots d\vecS_n .
\end{equation}
Note in particular that $\mu_{\Lambda}^{(n+1)}(\scrA\times\RR^d)=\mu_{\Lambda}^{(n)}(\scrA)$.

\section{A limiting random flight process}

In Theorem \ref{secThmMacro} we have identified a Markov process with memory two that describes the limiting distribution of billiard paths with random initial data $(\vecQ_0,\vecV_0)$. Let us denote by
\begin{equation}\label{SP}
	\{ \Xi(t) : t\in\RR_{>0} \},
\end{equation}
the continuous-time stochastic process that is obtained by moving with unit speed along the random paths $\vecS_1,\vecS_2,\ldots$ of the above Markov process with memory two. This random flight process is fully specified by the probability 
\begin{equation}\label{PProb-macro}
	\PP_{\Lambda}\big(\Xi(t_1)\in\scrD_1,\ldots, \Xi(t_M)\in\scrD_M\big)
\end{equation}
that $\Xi(t)$ visits the sets $\scrD_1,\ldots,\scrD_M\subset\T^1(\RR^d)$ at times $t=t_1,\ldots,t_M$, with $M$ arbitrarily large. To give a precise definition of \eqref{SP} set
$T_0:=0$, 
$T_n:=\sum_{j=1}^n \|\vecS_j\|$, and define the probability that $\Xi(t)$ is in the set $\scrD_1$ at time $t_1$ after exactly $n_1$ hits, in the set $\scrD_2$ at time $t_2$ after exactly $n_2$ hits, etc., by
\begin{multline} \label{PNLQ0LDEF}
	\PP_{\Lambda}^{(\vecn)}\big(\Xi(t_1)\in\scrD_1,\ldots, \Xi(t_M)\in\scrD_M, \; T_{n_1}\leq t_1< T_{n_1+1},\ldots, T_{n_M}\leq t_M< T_{n_M+1}  \big) \\
	:=
\mu_{\Lambda}^{(n+1)}\big(\big\{(\vecS_1,\ldots,\vecS_{n+1}): \Xi_{n_j}(t_j) \in\scrD_j,\; T_{n_j}\leq t_j< T_{n_j+1}\;(j=1,\ldots,M) \big\}\big) 
\end{multline}
with $\vecn:=(n_1,\ldots,n_M)$, $n:=\max(n_1,\ldots,n_M)$, and
\begin{equation}\label{Xn}
	\Xi_n(t):= \bigg( \sum_{j=1}^n \vecS_j + (t-T_n) \uvecS_{n+1}, \uvecS_{n+1}\bigg) .	
\end{equation}
The formal definition of \eqref{SP} is thus
\begin{multline} \label{PPLQ0LDEF}
		\PP_{\Lambda}\big(\Xi(t_1)\in\scrD_1,\ldots, \Xi(t_M)\in\scrD_M\big) \\ := \sum_{\vecn\in\ZZ_{\geq 0}^M} \PP_{\Lambda}^{(\vecn)}\big(\Xi(t_1)\in\scrD_1,\ldots, \Xi(t_M)\in\scrD_M \\ \text{ and } T_{n_1}\leq t_1< T_{n_1+1},\ldots, T_{n_M}\leq t_M< T_{n_M+1}  \big).
\end{multline}

The following theorem shows that the Lorentz process \eqref{LP} converges to the stochastic process \eqref{SP} as $\rho\to 0$. 

\begin{thm}\label{secThmMacro2}
Fix a lattice $\scrL$ and let $\Lambda$ be a Borel probability measure on $\T^1(\RR^d)$ which is absolutely continuous with respect to Lebesgue measure. Then, for any $t_1,\ldots,t_M\in\RR_{\geq 0}$, and any subsets $\scrD_1,\ldots,\scrD_M\subset \T^1(\RR^{d})$ with boundary of Lebesgue measure zero,
\begin{multline} \label{secThm-eq2-macro}
	\lim_{\rho\to 0}\Lambda\big(\big\{(\vecQ_0,\vecV_0)\in \T^1(\rho^{d-1}\scrK_\rho) :\\ (\vecQ(t_1),\vecV(t_1))\in \scrD_1,\ldots, (\vecQ(t_M),\vecV(t_M))\in \scrD_M \big\}\big) \\
= \PP_{\Lambda}\big(\Xi(t_1)\in\scrD_1,\ldots, \Xi(t_M)\in\scrD_M\big) .
\end{multline}
The convergence is uniform for $t_1,\ldots,t_M$ in compact subsets of $\RR_{\geq 0}$. 
\end{thm}

Theorem \ref{secThmMacro2} follows from Theorem \ref{secThmMacro}; the main ingredient in the proof is an estimate that shows that it is unlikely to have many collisions in any fixed time interval \cite[Section 5]{partII}.

The generalized linear Boltzmann equation \eqref{glB} can now be interpreted as the Fokker-Planck-Kolmogorov equation of the stochastic process \eqref{SP}, and its validity follows from Theorem \ref{secThmMacro2} by standard arguments from the theory of stochastic processes \cite[Section 6]{partII}.

\section{The distribution of free path lengths}

To give a detailed account of the proof of Theorem \ref{secThmMacro} would go beyond the scope of this lecture. I however hope to be able to explain the key idea, which is already clearly visible in the case $n=1$, i.e., the distribution of free path lengths.

In order to explain the proof, it will be more convenient to return to the original microscopic coordinates $(\vecq,\vecv)$, where the free path length diverges at the rate $\rho^{-(d-1)}$.

Let us denote by $\scrK_\rho\subset\RR^d$ the complement of the union of all scatterers.
The free path length for the initial condition $(\vecq,\vecv)\in\T^1(\scrK_\rho)$ is defined as
\begin{equation} \label{TAU1DEF0}
	\tau_1(\vecq,\vecv;\rho) = \inf\{ t>0 : \vecq+t\vecv \notin\scrK_\rho \}. 
\end{equation}
That is, $\tau_1(\vecq,\vecv;\rho)$ is the first time at which a particle with initial data $(\vecq,\vecv)$ hits a scatterer. We also include the (somewhat artificial) case when $\vecq\in\scrL$, i.e., when the particle starts at the center of a scatterer; in this case we think of $\scrK_\rho\subset\RR^d$ as the domain obtained by removing all scatterers {\em except} the one centered at $\vecq$. The following is the main result of \cite{partI}.

\begin{thm}\label{freeThm1}
Fix a lattice $\scrL$ of covolume one,
let $\vecq\in\RR^d$,
and let $\lambda$ be a Borel probability measure on $\S_1^{d-1}$ absolutely continuous with respect to Lebesgue measure. Then there exists a $\C^1$ function $F_{\scrL,\vecq}$ on $\RR_{\geq 0}$ such that, for every $\xi> 0$,
\begin{equation}
\lim_{\rho\to 0} \lambda(\{ \vecv\in\S_1^{d-1} : \rho^{d-1} 
\tau_1(\vecq,\vecv;\rho)\leq \xi \})
= F_{\scrL,\vecq}(\xi) .
\end{equation}
\end{thm}

The distribution of the free path lengths in the Lorentz gas was already investigated by Polya, who rephrased the problem in terms of the visibility in a (random and periodic) forest \cite{Polya18}. The problem of the limiting distribution in dimension $d=2$ was recently solved by  Boca and Zaharescu \cite{Boca07} in dimension $d=2$ in the case when $\vecq$ is either random or located at a lattice point; see also their earlier work with Gologan \cite{Boca03}, and the paper by Calglioti and Golse \cite{Caglioti03}. Previous work in higher dimension $d\geq 3$ includes the papers by Bourgain, Golse and Wennberg \cite{Bourgain98}, \cite{Golse00} who provide upper and lower bounds on the tail of the distribution of free path lengths. More details on the existing literature can be found in the survey \cite{Golse06}.

\section{The space of lattices}

The new idea in the joint work with Str\"ombergsson \cite{partI} is to translate the problem of the free path length into a question about the dynamics on the space of lattices. The advantage of this approach over previous attempts is that the technique extends naturally to arbitrary dimension, and that the limiting distributions have a canonical interpretation as the distribution function of random variables on a beautiful geometric object. 

A euclidean lattice $\scrL\subset \RR^d$ of covolume one can be written as $\scrL=\ZZ^d M$ for some $M\in\SLR$, where $\ZZ^d$ is the standard cubic lattice. Since $\SLZ$ leaves $\ZZ^d$ invariant, the homogeneous space $X_1=\SLSL$ parametrizes the space of lattices of covolume one. 
Similarly, let $\ASLR=\SLR\ltimes \RR^d$ be the semidirect product group with multiplication law
\begin{equation} \label{ASLMULTLAW}
	(M,\vecxi)(M',\vecxi')=(MM',\vecxi M' +\vecxi') .
\end{equation}
An action of $\ASLR$ on $\RR^d$ can be defined as
\begin{equation}
	\vecy \mapsto \vecy(M,\vecxi):=\vecy M+\vecxi .
\end{equation}
Each \textit{affine} lattice (i.e.\ translate of a lattice) 
of covolume one in $\RR^d$ can then be expressed as
$\ZZ^d g$ for some $g\in\ASLR$, 
and the space of affine lattices is then represented by $X=\ASLASL$ where $\ASLZ=\SLZ\ltimes \ZZ^d$.
We denote by $\mu_1$ and $\mu$ the Haar measure on $\SLR$ and $\ASLR$, respectively, normalized in such a way that they represent probability measures on $X_1$ and $X$. 

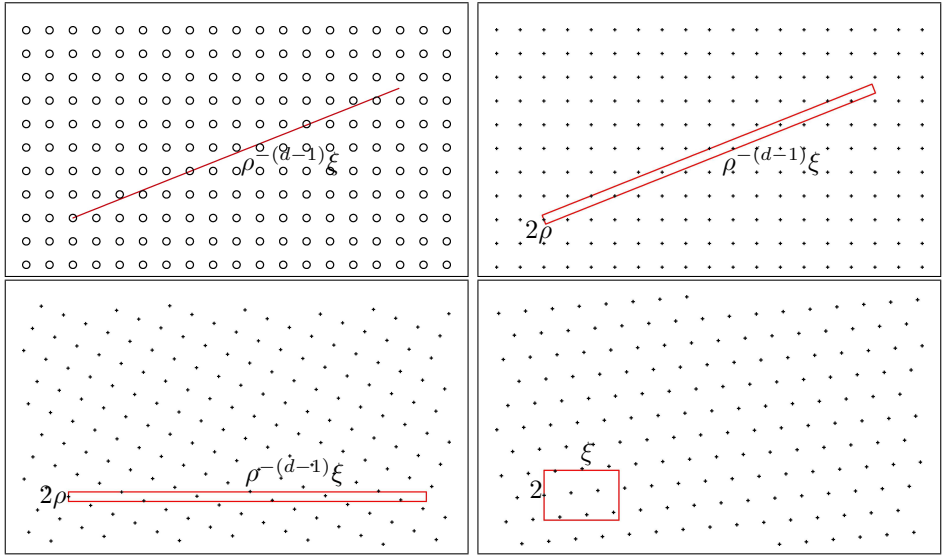
\begin{figure}
\begin{center}
\framebox{
\begin{minipage}{0.45\textwidth}
\unitlength0.1\textwidth
\begin{picture}(10,6)(0,0)
\put(0,0){\includegraphics[width=\textwidth]{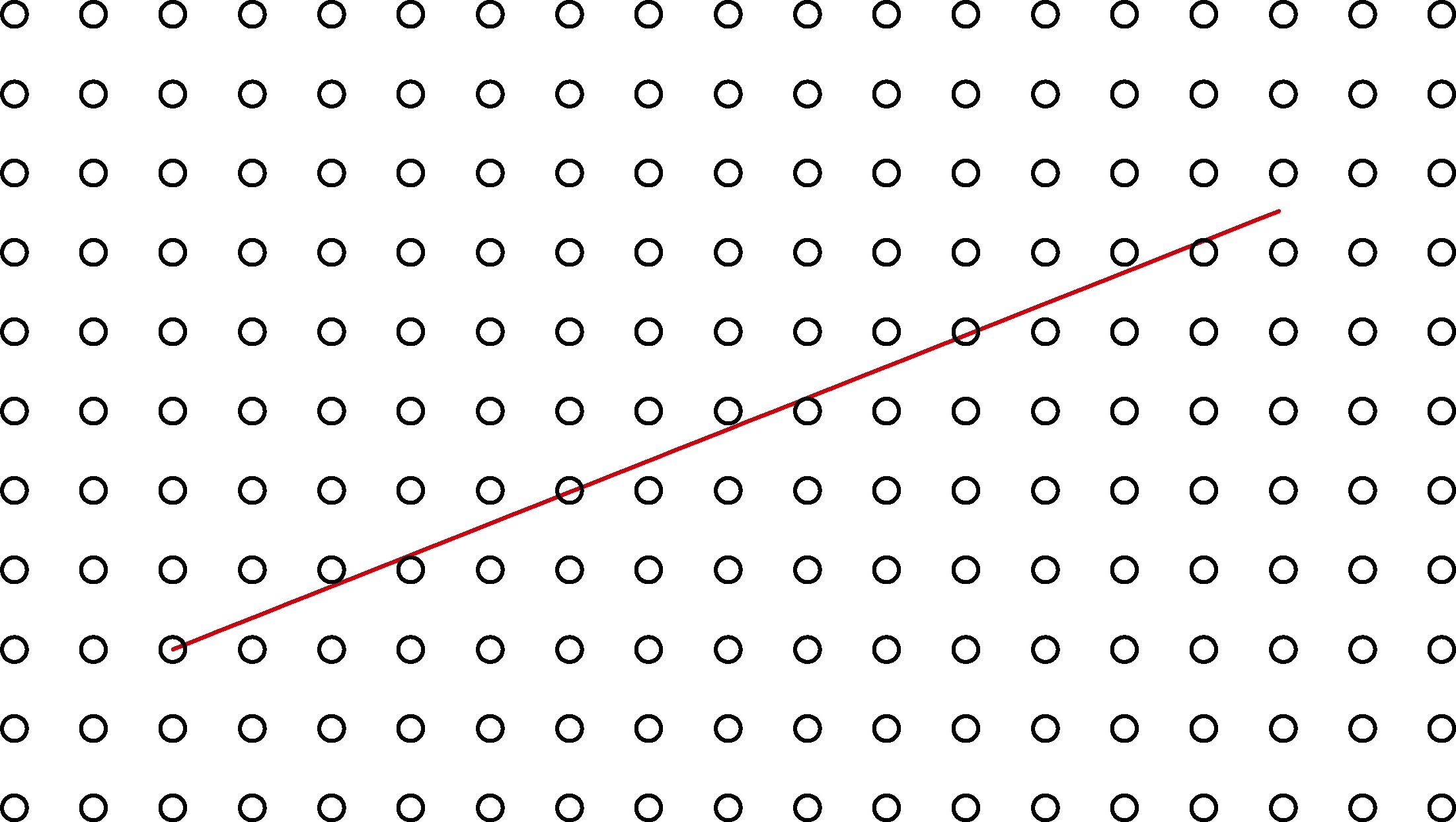}}
\put(5.1,2.3){$\rho^{-(d-1)}\xi$} 
\end{picture}
\end{minipage}
}
\framebox{
\begin{minipage}{0.45\textwidth}
\unitlength0.1\textwidth
\begin{picture}(10,6)(0,0)
\put(0,0){\includegraphics[width=\textwidth]{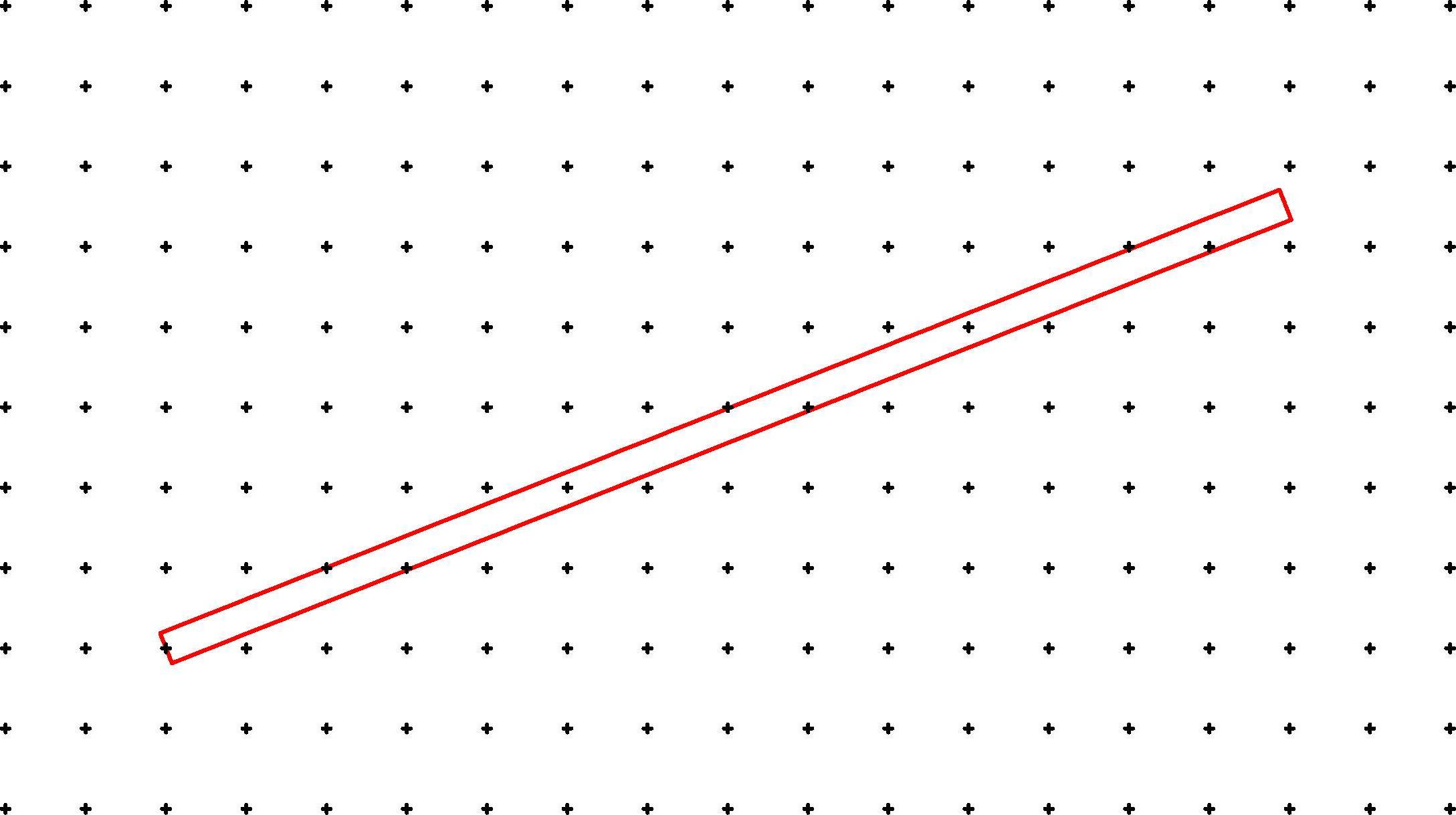}}
\put(0.7,0.7){$2\rho$} \put(5.3,2.3){$\rho^{-(d-1)}\xi$} 
\end{picture}
\end{minipage}
}
\framebox{
\begin{minipage}{0.45\textwidth}
\unitlength0.1\textwidth
\begin{picture}(10,6)(0,0)
\put(0,0){\includegraphics[width=\textwidth]{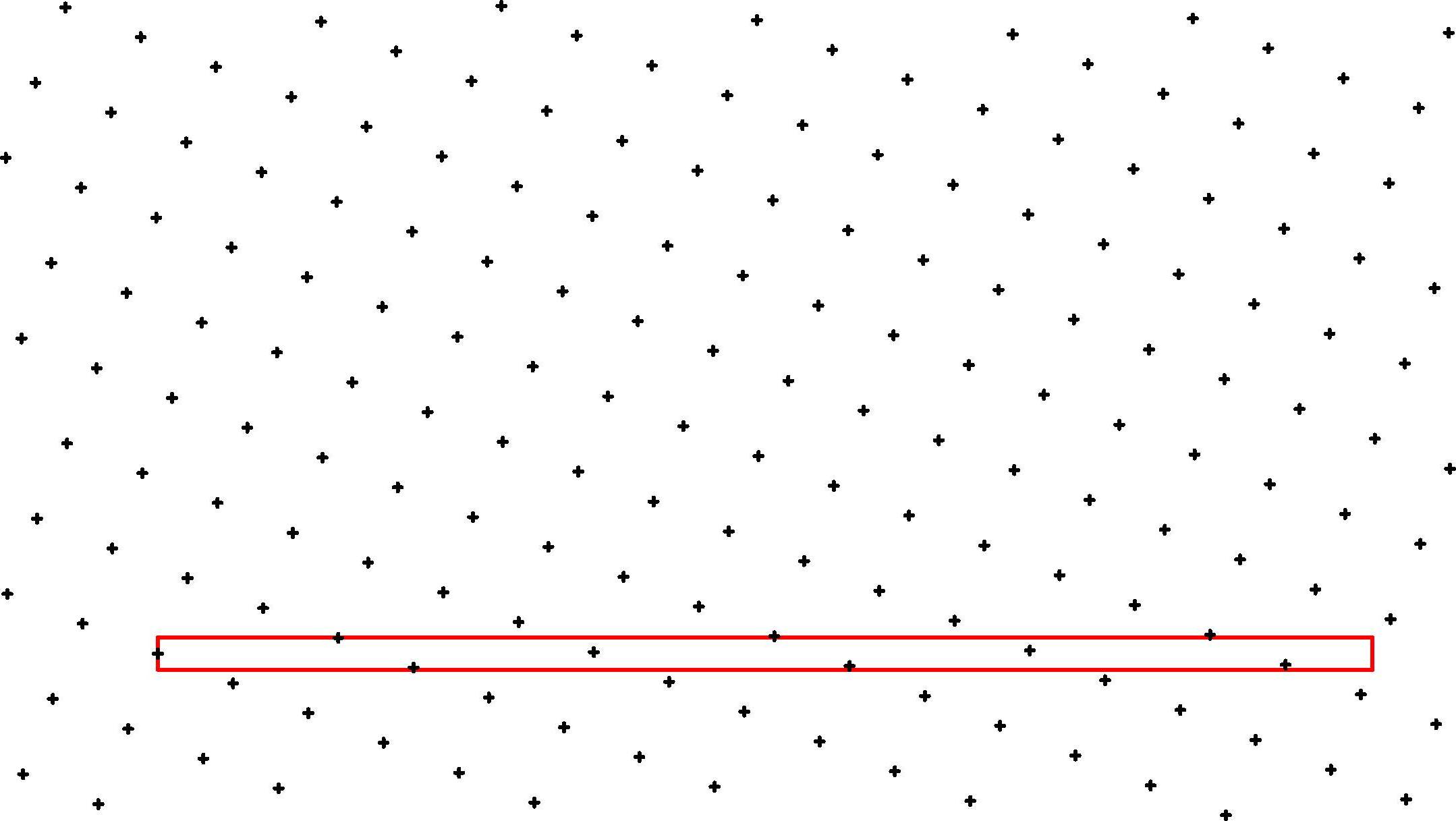}}
\put(0.4,1){$2\rho$} \put(5.2,1.5){$\rho^{-(d-1)}\xi$} 
\end{picture}
\end{minipage}
}
\framebox{
\begin{minipage}{0.45\textwidth}
\unitlength0.1\textwidth
\begin{picture}(10,6)(0,0)
\put(0,0){\includegraphics[width=\textwidth]{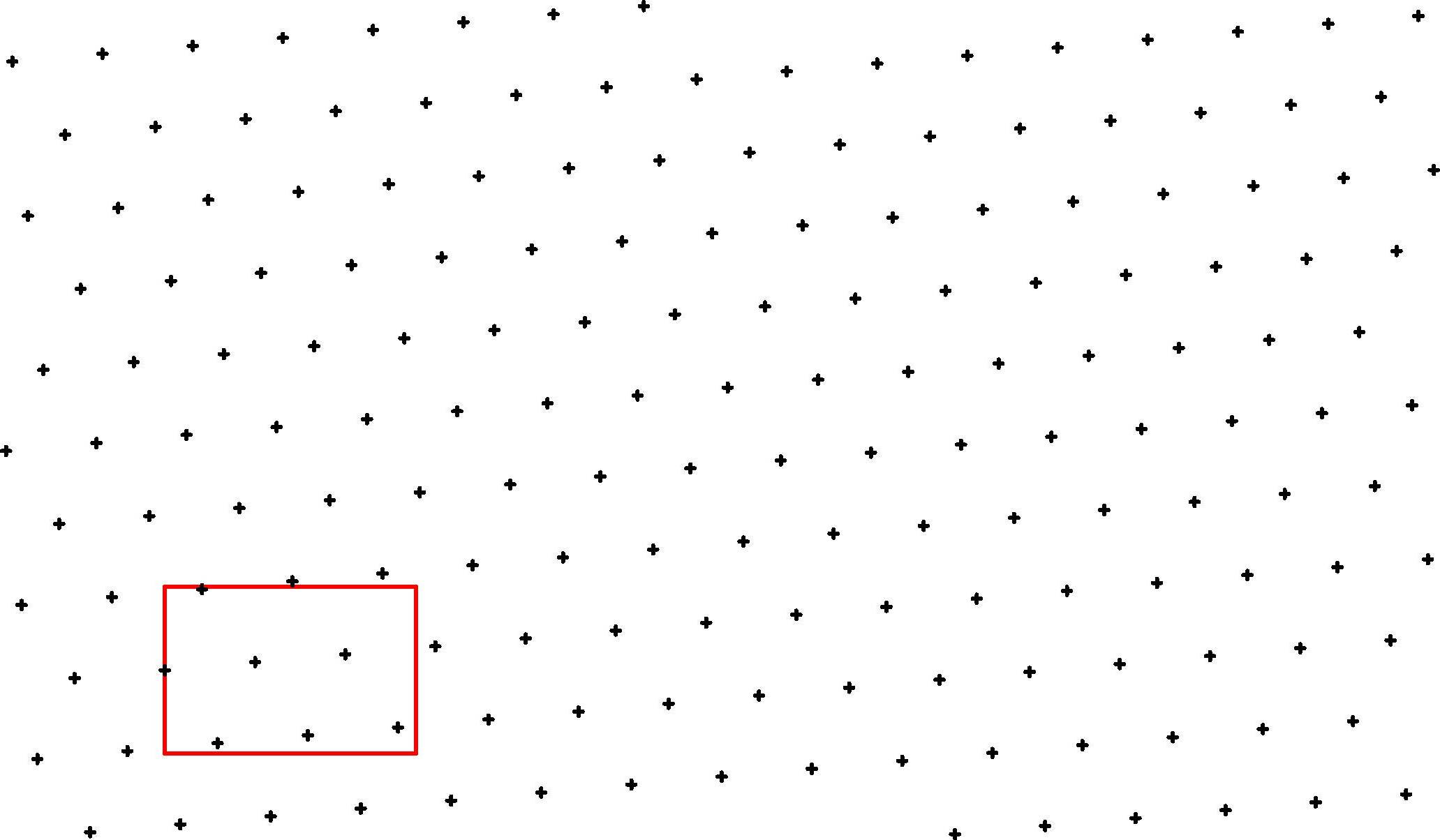}}
\put(0.8,1.1){$2$} \put(2,2){$\xi$} 
\end{picture}
\end{minipage}
}
\end{center}
\caption{Renormalization of the periodic Lorentz gas. {\em Step 1:} Replace the question on the number of intersections of a line with scatterers by a lattice point counting problem in a long stretched cylinder. {\em Step 2:} Rotate cylinder and lattice so the cylinder lies horizontally. {\em Step 3:} Apply a diagonal linear transformation that maps the cylinder to the $\rho$-independent reference cylinder $\scrZ(\xi)$. \label{figRN}}
\end{figure}

We are interested in the $\lambda$-measure of velocities with free path lengths at most $\rho^{-(d-1)}\xi$,
\begin{equation}
	\lambda(\{ \vecv\in\S_1^{d-1} : \rho^{d-1} 
\tau_1(\vecq,\vecv;\rho)\leq \xi \}).
\end{equation}
This is approximately the same as the $\lambda$-measure of directions such that a cylinder $\vecq+\scrZ(\vecv,\rho^{-(d-1)}\xi,\rho)$ of length $\rho^{-(d-1)}\xi$ and radius $\rho$, pointing in direction $\vecv$ contains at least one lattice point (cf.~Fig.~\ref{figRN}):
\begin{equation}
	\approx \lambda(\{ \vecv\in\S_1^{d-1} : \ZZ^d M \cap \vecq+\scrZ(\vecv,\rho^{-(d-1)}\xi,\rho) \neq \emptyset \}) .
\end{equation}
The approximation comes from the fact that our cylinder should have spherical caps of radius $\rho$ on each end; it is easy to show however that the $\lambda$-measure of $\vecv$ that have a lattice point in these caps is vanishingly small, as $\rho\to 0$ (see \cite[Section 4.1]{partI} for details). The next step is to shift lattice and cylinder by $-\vecq$, and then rotate by $K(\vecv)\in\SO(d)$ such that $K(\vecv)\vecv=\vece_1$, where $\vece_1=(1,0,\ldots,0)$: 
\begin{equation}
	=\lambda(\{ \vecv\in\S_1^{d-1} : (\ZZ^d M -\vecq) K(\vecv) \cap \scrZ(\vece_1,\rho^{-(d-1)}\xi,\rho) \neq \emptyset \}) .
\end{equation}
The cylinder now lies in the $\vece_1$-direction. We apply the linear transformation $D(\rho)=\diag(\rho^{d-1},\rho^{-1},\ldots,\rho^{-1})$ which transforms the long and thin cylinder into better proportions:
\begin{equation}\label{jjd}
	=\lambda(\{ \vecv\in\S_1^{d-1} : (\ZZ^d M -\vecq) K(\vecv)D(\rho) \cap \scrZ(\xi) \neq \emptyset \}) ,
\end{equation}
where
\begin{equation}
	\scrZ(\xi):=\scrZ(\vece_1,\xi,1)=\big\{(x_1,\ldots,x_d)\in\RR^d : 0 < x_1 <\xi, \|(x_2,\ldots,x_d)\|< 1 \big\}.
\end{equation}
Although the above linear transformations seem trivial, we have achieved a different perspective on the problem: Rather than counting lattice points in a long thin cylinder (which looks hard) we now count in a well proportioned object. The prize we have paid is that our original lattice $\scrL=\ZZ^d M$ has changed to $(\ZZ^d M -\vecq) K(\vecv)D(\rho)$. Hence we are moving through the space of (affine) lattices, as $\rho\to 0$, and may now employ ergodic theoretic methods to understand the averages over $\vecv$ with respect to $\lambda$. 

\section{Equidistribution in the space of lattices}

We begin with the simplest case, $\vecq\in\scrL$, i.e., without loss of generality $\vecq=\vecnull$. The right translation
\begin{equation}
	X_1 \to X_1, \qquad \SLZ M \mapsto \SLZ M \Phi^t
\end{equation}
by the element
\begin{equation}
	\Phi^t = \begin{pmatrix} \e^{-(d-1)t} & \vecnull \\ \trans\vecnull & \e^t 1_{d-1} \end{pmatrix}
\end{equation}
defines a flow on the homogeneous space $X_1=\SLSL$. This flow has many good chaotic features: it is ergodic, mixing and partially hyperbolic. 

The horospherical subgroups generated by
\begin{equation}
	n_+(\vecx)=\begin{pmatrix} 1_{d-1} & \trans\vecnull \\ \vecx & 1 \end{pmatrix},\qquad
	n_-(\vecx)=\begin{pmatrix} 1_{d-1} & \trans\vecx \\ \vecnull & 1 \end{pmatrix},
\end{equation}
respectively, generate the stable and unstable horospherical subgroups of the flow. Using the mixing property, a standard argument (see e.g. \cite{Eskin93}, \cite{Marklof00}) shows that averages over the unstable horosphere become asymptotically ($t\to\infty$) equidistributed in $X_1$ with respect to $\mu_1$.

\begin{thm}\label{thmh}
Let $\lambda$ be a Borel probability measure on $\RR^{d-1}$ absolutely continuous with respect to Lebesgue measure. Let $f:X_1\to\RR$ be bounded continuous, $M\in\SL(d,\RR)$. Then
\begin{equation}
	\lim_{t\to\infty} \int_{\RR^{d-1}} f(Mn_-(\vecx) \Phi^t) d\lambda(\vecv) = \int_{X_1} f(M') d\mu_1(M') .
\end{equation}
\end{thm}

Alternative proofs of this theorem can be obtained by using harmonic analysis (which is particularly feasible when $M=1$, i.e., the horosphere is closed) or Ratner's theory, which we will revisit below.

In view of \eqref{jjd} we are interested in the distribution of the orbit
\begin{equation}
	\SLZ \backslash \big\{ \SLZ  M K(\vecv) \Phi^t : \vecv\in\S_1^{d-1} \big\}
\end{equation}
in $X_1$, as $t\to\infty$ (set $t=\log 1/\rho$). By using the fact that this orbit is close to an unstable horosphere, we can exploit Theorem \ref{thmh} to deduce equidistribution also in this case; we refer the reader to \cite[Section 5]{partI} for details.

\begin{thm}\label{thmr}
Let $\lambda$ be a Borel probability measure on $\S_1^{d-1}$ absolutely continuous with respect to Lebesgue measure. Let $f:X_1\to\RR$ be bounded continuous, $M\in\SL(d,\RR)$. Then
\begin{equation}
	\lim_{t\to\infty} \int_{\S_1^{d-1}} f(M K(\vecv)\Phi^t) d\lambda(\vecv) = \int_{X_1} f(M') d\mu(M') .
\end{equation}
\end{thm}

This theorem thus states, that in the limit $t\to\infty$ we can replace the $\lambda$-average over $\vecv$ by an average of the entire space of lattices. This yields (modulo some technicalities) the proof of the limit law for the free path length, Theorem \ref{freeThm1}, in the case $\vecq\in\scrL$, plus a formula for the limit distribution:
\begin{equation}\label{eqF0}
	F_{\scrL,\vecnull}(\xi) =
	\mu_1(\{ M\in X_1: \ZZ^d M \cap \scrZ(\xi) \neq \emptyset \}) .
\end{equation}
That is, {\em the limit distribution of the free path length for a particle emerging from the center of a scatterer is given by the probability that a random lattice intersects the cylinder $\scrZ(\xi)$ in at least one point.} Note that the limit distribution $F_\vecnull(\xi):=F_{\scrL,\vecnull}(\xi)$ is independent of the lattice $\scrL$ and of $\lambda$.

Instead of particles emerging from a lattice point we can also consider initial conditions on the boundary of a scatterer; this leads to different limit distributions and is one of the crucial steps in the proof of the existence of the limiting random flight process described in Theorems \ref{secThmMacro} and \ref{secThmMacro2}.

Let us now turn to the case $\vecq\not\in \QQ\scrL$. 
The right translation
\begin{equation}
	X \to X, \qquad \ASLZ g \mapsto \ASLZ g \Phi^t
\end{equation}
by the element
\begin{equation}
	\Phi^t = \left(\begin{pmatrix} \e^{-(d-1)t} & \vecnull \\ \trans\vecnull & \e^t 1_{d-1} \end{pmatrix}, \vecnull \right) 
\end{equation}
now defines a flow on the homogeneous space $X=\ASLASL$. In analogy with the above, we set
\begin{equation}
	n_-(\vecx)=\left(\begin{pmatrix} 1_{d-1} & \trans\vecx \\ \vecnull & 1 \end{pmatrix}, \vecnull \right) .
\end{equation}
The crucial difference is now that $n_-(\vecx)$ no longer generates the full unstable horosphere for the flow $\Phi^t$, and hence the mixing argument is no longer sufficient. Instead, we need to employ Ratner's classification of measures that are invariant under unipotent actions, which in the present case is given by the right action of $n_-(\vecx)$. We can in particular exploit a very useful theorem of Shah \cite{Shah96} to show the following. (See \cite[Section 5]{partI} for details, and \cite{nato} for a general introduction to applications of Ratner's theory to problems of this kind.)

\begin{thm}
Let $\lambda$ be a Borel probability measure on $\RR^{d-1}$ absolutely continuous with respect to Lebesgue measure. Let $f:X\to\RR$ be bounded continuous, $\vecalf\in\RR^d\setminus\QQ^d$, $M\in\SL(d,\RR)$. Then
\begin{equation}
	\lim_{t\to\infty} \int_{\RR^{d-1}} f((1,\vecalf)(M,\vecnull)n_-(\vecx) \Phi^t) d\lambda(\vecx) = \int_{X} f(g) d\mu(g) .
\end{equation}
\end{thm}

Using the same approximation argument that led to Theorem \ref{thmr} we deduce:

\begin{thm}
Let $\lambda$ be a Borel probability measure on $\S_1^{d-1}$ absolutely continuous with respect to Lebesgue measure. Let $f:X\to\RR$ be bounded continuous, $\vecalf\in\RR^d\setminus\QQ^d$, $M\in\SL(d,\RR)$. Then
\begin{equation}
	\lim_{t\to\infty} \int_{\S_1^{d-1}} f((1,\vecalf)(M,\vecnull)(K(\vecv),\vecnull) \Phi^t) d\lambda(\vecv) = \int_{X} f(g) d\mu(g) .
\end{equation}
\end{thm}

This proves Theorem \ref{freeThm1} in the case $\vecq\notin\QQ\scrL$, and yields the formula
\begin{equation}\label{eqF}
	F_{\scrL,\vecq}(\xi) =
	\mu(\{ (M,\vecx)\in X: (\ZZ^d M +\vecx) \cap \scrZ(\xi) \neq \emptyset \}) .
\end{equation}
Hence, {\em the limit distribution of the free path length for a particle starting at a generic position is given by the probability that a random affine lattice intersects the cylinder $\scrZ(\xi)$ in at least one point.}
Here, $F(\xi):=F_{\scrL,\vecq}(\xi)$ is evidently independent of $\scrL$, $\vecq$ and $\lambda$.

I omit the discussion of the remaining $\vecq\in \QQ\scrL$---the arguments are analogous to the above \cite{partI}.

Only in dimension $d=2$ we have been able to turn the formulas \eqref{eqF0} and \eqref{eqF} for $F_\vecnull(\xi)$ and $F(\xi)$ into explicit functions of $\xi$. The easiest approach is to integrate our formula for the transition probability \eqref{Xp} (see \cite{partIII} for details),
\begin{equation}
	F_\vecnull(\xi) = \int_0^\xi \int_{-1}^1 \varPhi_\vecnull(\xi',0,z)\, dz\,d\xi',
\end{equation}
\begin{equation}
	F(\xi) = \int_0^\xi \int_{\xi'}^\infty \int_{-1}^1 \int_{-1}^1  \varPhi_\vecnull(\xi'',w,z)\, dw\, dz\,d\xi''\,d\xi' .
\end{equation}
The resulting explicit expressions coincide with those obtained, using different methods, by Boca, Gologan and Zaharescu \cite{Boca03}, and by Boca and Zaharescu \cite{Boca07}, respectively.

\section{Asymptotics}

The geometry of the space of lattices is significantly more complicated in dimension $d\geq 3$, and it seems extremely hard to obtain any explicit formulas. It is possible, however, to describe the asymptotic tails of our limit distributions, by observing that when $\xi>0$ is very large (or small), then the lattices that contribute to $F_{\vecnull}(\xi)$ and $F(\xi)$ must have at least one very short basis vector. That is, all integration is restricted to the cusps of the spaces $X_1$ and $X$, respectively, whose geometry is simpler. Using tools from the geometry of numbers we can show that \cite{partIV}
\begin{equation}
	F_{\vecnull}(\xi)=1 \qquad \text{for $\xi$ sufficiently large},
\end{equation}
and
\begin{equation}
	F_{\vecnull}(\xi)=\frac{\nu_d}{\zeta(d)}\;\xi
+O(\xi^2), \qquad \xi\to 0,
\end{equation}
where $\nu_d=\frac{\pi^{(d-1)/2}}{\Gamma((d+1)/2)}$ is the volume of the $(d-1)$-dimensional unit ball.
Similarly,
\begin{equation}
	F(\xi)=1
-\frac{\pi^{\frac{d-1}2}}{2^{d}d\, \Gamma(\frac {d+3}2)\,\zeta(d)}\;\xi^{-1}
+O\bigl(\xi^{-1-\frac 2d}\bigr),
\qquad \xi\to\infty ,
\end{equation}
and
\begin{equation}
	F(\xi)=\nu_d\; \xi+O\bigl(\xi^2\bigr) , \qquad \xi\to 0 .
\end{equation}
We also obtain asymptotic formulas for the collision kernel $\varPhi_{\vecnull}(\xi,w,z)$ in the limits $\xi\to 0$ and $\xi\to\infty$, see \cite{partIV} for details.

\section{Outlook}

The techniques outlined above do not necessarily require that the scatterers are rigid spheres. It is sufficient to assume that the scattering map is dispersive; a Muffin-Tin Coulomb potential would be a good example \cite{partII}. A key hypothesis of our approach is however is that the interaction region of the scattering process is finite, so that the test particle moves along straight lines for most of the time. This assumption is no longer valid in the case of long-range potentials. Provided the renormalization approach can be modified accordingly, it seems feasible to generalize our studies to crystals with long-range potential, at least for sufficiently fast power-like decay. A result by Desvillettes and Pulvirenti \cite{Desvillettes99} achieves this objective (with some additional technical assumptions) in the case of random scatterer configurations; cf. also the work by Poupaud and Vasseur \cite{Poupaud03}.

A related problem is to consider different scaling limits for compact potentials, where the strength of the potential is reduced, and at the same time the scatterer density rescaled suitably to achieve a non-trivial limit. In this case grazing collisions become important, and one expects a different kinetic equation for the macroscopic dynamics; cf.~Desvillettes and Ricci \cite{Desvillettes01} for the corresponding result for a random scatterer configuration---here the limiting kinetic equation is the classical Fokker-Planck equation.

The renormalization approach we have developed for the Lorentz gas assumes that the scattering process is the same for each scatterer. It is possible to remove this assumption, as long as the resulting (global) potential of the crystal is still periodic, or in the case of quasicrystals. A natural question leading on from this is whether one can describe the kinetic equations in the case of a (quasi-)crystal with random defects, i.e., a scatterer is removed from a lattice site with probability $p$, with $0< p < 1$. Such a set-up will lead to an interesting variation of the renormalization method, since the modular invariance of the crystal will be replaced by a modular invariance in distribution (we assume $p$ is the same for each scatterer). It is no surprise that the limiting case $p\to 1$ leads back to the linear Boltzmann equation \cite{Ricci04}. 

A further important case that has been extensively studied for stochastic Lorentz gases is the dynamics in the presence of electro-magnetic fields \cite{Desvillettes04}. In the case of constant magnetic or electric fields, the above problem has some connection with beautiful, basic number-theoretic questions on the distribution of lattice points near circles or parabolas, respectively. 

Finally, an obvious challenge is to adapt our renormalization approach to the quantum mechanical problem, and derive a corresponding quantum kinetic equation, in suitable weak-coupling or low-density limits discussed for random scatterer configurations by  Erd\"os and Yau \cite{Erdos00}, and Eng and Erd\"os \cite{Eng05}.
It should be stressed that of course the quantum theory of electrons in {\em fixed} periodic potentials is well understood, and that the lattice symmetry allows for a wealth of techniques in the spectral analysis (Floquet-Bloch decomposition). In particular the KKR method is an extremely useful tool, and has been applied to the semiclassical analysis of the symmetry-reduced periodic Lorentz gas (the Sinai billiard) in connection with quantum chaos, see Berry \cite{Berry81}, and (for the case of small scatterers) Dahlqvist and Vattay \cite{Dahlqvist98}. 

The central idea of the work presented here is to exploit the dynamics of flows on the space of lattices. This technique has proved to be extremely powerful also in other applications in mathematical physics, including KAM theory \cite{Khanin06}, \cite{Khanin07}, arithmetic quantum unique ergodicity \cite{Lindenstrauss06} and the Berry-Tabor conjecture \cite{Eskin05}, \cite{Marklof98}, \cite{pairI}, \cite{pairII}. Flows on the space of lattices (and more general homogeneous spaces) should be viewed as {\em the} higher-dimensional generalization of the classical continued fraction algorithms---I expect more striking applications in the future.

\section*{Acknowledgements}

All of the results described above have been obtained in collaboration with Andreas Str\"ombergsson. It has been a great pleasure to work with him on this project. I would also like to thank EPSRC, the Leverhulme Trust, the Royal Society and the Wolfson Foundation for their generous support.

\end{document}